\documentstyle[preprint,aps]{revtex}
\tightenlines
\begin{document}
\title{Wave localization in generalized Thue-Morse superlattices with disorder}
\author{Lev I. Deych, D. Zaslavsky, and A.A. Lisyansky}
\address{Department of Physics, 
Queens College of City University of New York, Flushing, NY 11367}
\date{\today}
\maketitle

\begin{abstract}
In order to study an influence of correlations on the localization
properties of classical waves in random superlattices we introduce a
generalized random Thue-Morse model as a four-state Markov process with two
parameters that determine probabilities of different configurations. It is
shown that correlations can bring about a considerable change in the
transmission properties of the structures and in the localization
characteristics of states at different frequencies.
\end{abstract}
\pacs{42.25.Bs,03.40.Kf,41.20.Jb}
\section{Introduction}

The effects of correlation on localization properties of electrons and
classical waves in one dimensional disordered systems has recently attracted
a great deal of attention. For the canonical Anderson model \cite{Anderson}
with uncorrelated diagonal disorder, it is a well established fact that
almost all states in 1-$D$ systems are localized, ensuring the absence of
transport through such systems. Correlation between, for example, random
values of energy at different sites was proven to change this situation
dramatically. This was shown for the first time in Ref.\cite{RDM}, where the
so called random dimer model was introduced. In this model, the same value
of energy was randomly assigned to pairs of consecutive sites, which
introduced ``rigid'' correlations between energies at consecutive sites. It
was shown that in such a model $\sqrt{N}$, where $N$ is a number of sites,
states remain delocalized. These delocalized states appear in the vicinity
of certain resonant values of energy. The random dimer model is in some
aspects analogous to classical wave propagation through a random
superlattice constructed from different layers with fixed thickness stacked
at random. It was shown in Ref.\cite
{randommedia,randomsuperlattice1,randomsuperlattice2} that in the
superlattice with two randomly positioned layers there exist two resonance
frequencies at which the transmissivity of the system becomes equal to one.
In both random dimer and random superlattice models, the dimers or layers
themselves were assumed to be distributed randomly without correlation. It
is interesting, however, to study how some additional correlations between
different blocks of these models affects the localization properties. For
the dimer model this question was addressed in Ref.\cite
{thermalcorrelations,randommatrix1,randommatrix2}. The first of these papers
dealt with the effects of thermally induced correlations on the localization
length of a random dimer harmonic chain. In Ref.\cite{randommatrix1}, a
dependence of the localization length upon the correlation radius of a
Markov sequence of the product of random matrices was studied, and
fluctuations of the Lyapunov exponent in the system of finite size were
investigated in Ref.\cite{randommatrix2}.

The transmission coefficient and localization length of acoustic waves in
random correlated superlattices were considered in Ref.\cite{correlsuperlat}
. Correlations in the latter paper were introduced by constructing the
superlattice according to three different Markov processes. The
Hendricks-Teller model (HT), the randomized Markov versions of Fibonacci,
and Thue-Morse (TM) sequences were considered. The Hendricks-Teller model is
a version of a dichotomous process, which is known to result in stochastic
structure with an exponential correlation function. The main feature of
Fibonacci and TM superlattices compared to the HT model is the presence of
short-range order. It was found in Ref.\cite{correlsuperlat} that the
frequency dependence of the transmission coefficient is quantitatively
different for the first two models and the last one. One can assume that
this difference is due to the difference in short-range structure of the
systems.

In the present paper, we proceed with a detailed study of the effects of the
short-range correlations on the localization properties of 1-$D$ random
systems. For the sake of concreteness, we deal with scalar wave propagation
through a random superlattice. Our results, however, can be applied to dimer
models as well. We consider a random superlattice constructed from two
layers $A$ and $B$ with different characteristics (dielectric constants, for
instance, in the case of electromagnetic wave propagation) stacked at random
according to the rules described in the first section of the paper. These
rules introduce a generalized Markov Thue-Morse model. This model can be
reduced to the canonical random TM model considered in Ref.\cite
{correlsuperlat} by selecting proper probabilities. Our model can also be
reduced to the HT model with exponential correlations, so we will be able to
investigate an interplay between ``soft'' exponential correlations and a
more ``rigid'' short-range order introduced by Thue-Morse-like rules.

\section{The model and its statistical properties.}

We consider the propagation of classical waves through one-dimensional
random media. This model corresponds to propagation of elastic or
electromagnetic waves through a layered medium which is random in the
direction of propagation of waves and homogeneous in the transversal
direction. For the case of normal incidence, the vector nature of waves can
be neglected since no conversion between different polarizations occurs and
one can consider the scalar wave equation: 
\begin{equation}
\frac{d^{2}E}{dx^{2}}+k_{0}^{2}\epsilon (x)=0,  \label{waveeq}
\end{equation}
where $k_{0}=\omega /c$ is the wave vector of the wave with the frequency $
\omega $ propagating with speed $c$ in a homogeneous medium surrounding a
disordered material. The parameter $\epsilon (x)$ describes a superlattice
composed of two different layers with the same thickness $d$, so that $
\epsilon (x)$ takes two different values $\epsilon _{1}$ and $\epsilon _{2}$
for each of the layers. These layers are stacked together at random
according to the following rules. If a layer is the first one in a sequence
of similar layers then the probability for the second like layer to appear
is $p$. If two like layers already appear in a sequence the probability for
the third consecutive like layer to occur is equal to $q$. These rules
introduce a four-state Markov process with the following conditional
probabilities: $P(AB\mid B)=P(BA\mid A)=p$; $P(AA\mid A)=P(BB\mid B)=q$. The
conventional TM Markov superlattice considered in \cite{correlsuperlat}
corresponds to $p=1/2,q=0$. This choice of parameters $p$ and $q$
``forbids'' the occurrence of blocks of the same layers with a length of
more than two. Another interesting realization of this model which in a
sense is opposite to the TM model, arises if one takes $p=1$ and $q=1/2$. In
this case blocks with the length less than $2$ are forbidden. One should not
confuse, however, this case with a simple dimerization of layers. In the
latter case only blocks with even numbers of like layers can appear, which
is obviously equivalent to doubling of layers' thicknesses. In the model
proposed here blocks with odd number of like layers and blocks with even
numbers of layers can occur. For $p=q$ the model reproduces the properties
of the so called dichotomous process (two-state Markov process) with $p$
being the transition probability from one state to another. This is proven
to result in a exponential correlation function with the correlation length $
l_{exp}=(-\ln |2p-1|)^{-1})$. Hence, we can conclude that the proposed model
exhibits two kinds of correlations. First, we have short-range correlations
determined by the specific short- range order presumed in the model with
fixed correlation length $l_{sh}=2d$, and with the degree of correlation
being proportional to the difference $\mid p-q\mid $. For $q\ne 1/2$ we have
exponentially decreasing correlations of the HT kind. Figs. 1(a-c) present
the results of numerical calculations of the correlation function$
K(r_{1}- r_{2})=\langle \epsilon (r_{1})\epsilon (r_{2})\rangle -\langle
(\epsilon )^{2}\rangle $, where angle brackets denote averaging over
different realizations of the random function $\epsilon (r)$. For numerical
averaging we use 10,000 realizations of a superlattice constructed
in accord with the rules described above. Calculations were carried out for
superlattices of different lengths and with different choices of the
starting point. We found that correlation functions do not depend upon the
size of the system or upon the starting point. The inserts in these figures present the Fourier transforms $S(q)$ of the corresponding correlation functions. These results demonstrate
that the system indeed has both short-range and long-range correlations.\footnote{We use the term ``long-range correlations" to refer to {\it exponential} correlations, which can have a correlation radius larger than the  layer's thikness.} Short-range
correlations at the scale of several thicknesses of a layer cause oscillations of the correlation functions and corresponding maxima on their Fourier transforms. In the system with  short-range correlations only the function $S(q)$ takes zero value at $q=0$ [Fig. 1(a)]. These correlations disappear when $\mid p-q\mid\rightarrow 0$.
Long-range correlations exist  for $q\ne 1/2$ only and are responisble for the exponential tail of the correlation functions. They reduce the maximum of the function $S(q)$  and cause its smoother decrease  for larger $q$ [Fig. 1(b)].  At $q<1/2$ these correlations are actually ``anticorrelations''
since they favor the appearance of of different blocks at adjacent
positions. Two extreme cases with $p=1$, $q=0$ and $p=0$, $q=1$ correspond
to fully ordered periodic structures with periods $2d$ and $d$,
respectively. The case $q=1$ results in homogeneous structures consisting of
one type of block only. It can be either block $A$ or $B,$ whichever block
occurs first.

Localization properties of a wave propagating through a random superlattice
are determined by scattering from interfaces between blocks of the same
layers. Therefore, an important statistical characteristic of the system
relevant to wave propagation is the distribution of block lengths $P(n),$
where $n$ is the number of layers of the same kind constituting a block, and 
$P$ is the probability of finding a block of length $n$. For our model, this
function can be shown to be 
\begin{eqnarray}
&&P(1)=1-p  \nonumber \\
&&P(n)=p(1-q)q^{n-2\text{ \ \ }}\text{for }n\ge 2  \label{widthdistribution}
\end{eqnarray}
In the extreme case $p=1$, $q=1/2$ ,Eq. (\ref{widthdistribution}) takes the
form $P(1)=0$, $P(n)=(1/2)^{n-1}$, $n>1$, which is quite similar to the
result for an uncorrelated superlattice $P(n)=(1/2)^{n}$. We show, however,
that a seemingly small discrepancy between these two distributions results
in a considerable qualitative difference between localization properties of
waves propagating in corresponding media. The average lengths of $A$ and $B$
blocks $\langle D_{A,B}\rangle $ are equal to each other, 
\begin{equation}
\langle D_{A}\rangle =\langle D_{B}\rangle =d\frac{1+p-q}{1-q},
\label{averagewidth}
\end{equation}
where the total length of the system is assumed to be infinite. This
expression diverges at $q\rightarrow 1$ which merely reflects the fact that
at $q=1$ the entire superlattice is composed of the same blocks, so the
average length of this block is equal to the total length of the system
assumed to be infinite.

\section{Average transmission rate and Lyapunov exponent}

In order to simulate wave propagation through the system, we make use of the
transfer-matrix method. A transfer-matrix connects an amplitude $E_{n}$ and
its first derivative $E_{n}^{\prime }=dE/dx$ of the field in $n$\/th slab
with the corresponding values in the $(n+1)$th slab: 
\begin{equation}
{\bf u}_{n+1}={\bf T}_{n}{\bf u}_{n},  \label{dynamics}
\end{equation}
where ${\bf u}_{n}$ is a vector with components $E_{n}$ and $E_{n}^{\prime }$
, and ${\bf T}_{n}$ is the transfer-matrix determined as follows: 
\begin{equation}
{\bf T}_{n}=\left( 
\begin{array}{cc}
\cos (k_{n}d) & \frac{1}{k_{n}}\sin (k_{n}d) \\ 
-k_{n}\sin (k_{n}d) & \cos (k_{n}d)
\end{array}
\right) ,  \label{Tmatrix}
\end{equation}
where $k_{n}=k_{0}\sqrt{\epsilon _{n}}$ is a wave number in $n$\/th layer.
The transmission coefficient $T$ is determined by the equation: 
\[
{\bf u}_{N}=\hat{{\bf T}}_{N}\text{\/\thinspace }{\bf u}_{0},
\]
where 
\[
{\bf u}_{N}=\left( 
\begin{array}{c}
t \\ 
itk_{0}
\end{array}
\right) 
\]
describes a wave transmitted through the superlattice, and 
\[
{\bf u}_{0}=\left( 
\begin{array}{c}
1+r \\ 
-irk_{0}
\end{array}
\right) 
\]
corresponds to incident and reflected waves. The transmission rate is
defined according to $T=\mid t\mid ^{2}$. The matrix ${\hat{{\bf T}}}_{N}$
is the product of all $T$-matrices corresponding to each layer: 
\[
{\hat{{\bf T}}}_{N}=\prod_{1}^{N}{\bf T}_{i}.
\]
The Lyapunov exponent $\gamma $ is determined according to 
\begin{equation}
\gamma =\lim_{N\rightarrow \infty }\ln {\hat{{\bf T}}}_{N}  \label{Lyapunov}
\end{equation}
and is known to be a self-averaging quantity in the limit of an infinite
system. For a system of a finite size this is a random variable. To
characterize the statistical properties of the Lyapunov exponent one can use
its statistical momenta, such as a mean value, mean-root-square
fluctuations, ets. \cite{randomsuperlattice2}. Another approach exploited in
Ref. \cite{randommatrix2} uses the generalized Lyapunov exponent.

Our analytical calculations utilize the approach developed in Ref. \cite
{randomsuperlattice2}. The authors of this work established a useful
relationship between the backscattering rate of waves , $\ell (L,\omega )$,
and the structure factor of a superlattice, $I_{s}(\omega )$: 
\begin{equation}
\ell (L,\omega )=\frac{2R^{2}}{\langle D_{A}\rangle +\langle D_{B}\rangle }
I_{s}(L,\omega ).  \label{backscattering}
\end{equation}

The structure factor in an infinite system is given by the following
expression 
\begin{equation}
I_{s}(\omega )=Re\left[ \frac{(1-\varepsilon _{A})(1-\varepsilon _{B})}{
1-\varepsilon _{A}\varepsilon _{B}}\right] ,  \label{structure}
\end{equation}
where $\varepsilon _{A}=\langle \exp (-2idk_{A}n_{A})\rangle $ and $
\varepsilon _{B}=\langle \exp (-2idk_{B}n_{B})\rangle $ and averaging is
carried out over the distribution $P(n)$ of thicknesses of corresponding
blocks given by Eq. (\ref{widthdistribution}). It is straightforward to show
that for this distribution 
\begin{equation}
\varepsilon _{j}=\exp (-2idk_{j})\frac{1-p+(p-q)\exp (-2idk_{j})}{1-q\exp
(-2idk_{j})},  \label{averageepsilon}
\end{equation}
where $j=A,B$. Frequencies, for which $2dk_{j}=2\pi n,n=0,1,2...$,
correspond to the resonant transmission with $T=1$ in a system of any size.
These frequencies are present in a system with any type of statistical
distribution of layers, therefore we will call them fundamental resonances.
The short-range correlations, which occur when $p\ne q$, bring about new
characteristic frequencies associated with the term $\exp (-4idk_{j})$. We
will see later that these frequencies actually manifest themselves as some
additional maxima on frequency dependences of the localization length and
the transmissivity. The effect of this term is the most prominent for $p=1$.
The expression for $I_{s}(\omega )$ in the general case is rather cumbersome
so that we only show it for the special case $p=1$, $q=1/2$: 
\begin{equation}
I_{s}=\frac{4(1-\cos 2k_{A}d)(1-\cos 2k_{B}d)(5+2\cos 2k_{A}d+2\cos 2k_{B}d)
}{\mid 4-2\exp (-2\imath k_{A}d)-2\exp (-2\imath k_{B}d)+\exp [-2\imath
(k_{A}+k_{B})d]-\exp [-4\imath (k_{A}+k_{B})d]\mid ^{2}}.  \label{Is}
\end{equation}

The backscattering length $\ell (\omega )$ was shown in Ref.\cite
{randomsuperlattice2} to determine the Lyapunov coefficient, $\gamma (\omega
)$: 
\begin{equation}
\gamma =\frac{1}{2\ell (\omega )}.  \label{lyapunov}
\end{equation}
With the localization length found one can calculate the average
transmission rate, fluctuations of the transmission and other relevant
characteristics \cite{randomsuperlattice1,randomsuperlattice2}. In Figs. 2
and 3 we present results of numerical and analytical calculations for the
average transmission coefficient and the Lyapunov exponent for different
kinds of superlattice. For simulations we used superlattices with 300 layers
and the ratio between layer parameters $\epsilon _{A}/\epsilon _{B}=1.2$. We
averaged over 500 different randomly chosen realizations of the system.
Figs. 2(a) and 2(b) show the average transmission for uncorrelated and HT
superlattices, respectively. They reproduce results of Ref.\cite
{randomsuperlattice1,randomsuperlattice2,correlsuperlat}. Fig. 2(c) presents
the frequency dependence of the average transmission for our model with
``rigid'' short-range correlation $p=1$, and with no exponential correlation 
$q=1/2$. One can see that the average transmission reacts sharply on the
short-range order: new maxima appear between fundamental resonance
frequencies. The TM Markov superlattice also results in some structure in
the frequency dependence of the average transmission rate \cite
{correlsuperlat}. However, the magnitude of the transmission at these new
maxima for TM case is negligible for the lattice compounded of 300 layers.
The authors the Ref.\cite{correlsuperlat} used a system with only 64 layers
in their calculations. Therefore in order to compare effects of different
kinds of short-range order we show in Fig. 2(d) results of calculations of
average transmission coefficient for the TM superlattice and our model with $
p=1$, $q=1/2$ for the system with 64 layers. This drastic decrease in
transmission rate for TM model is obviously due to the sharp increase of
scattering interfaces in it compared to our situation.

Fig. 2(e) presents the average transmission for the case when both
short-range and exponential correlations are present ($p=1$, $q=0.8$). Such
correlations favor like blocks stacked together, therefore we observe an
overall increase of the average transmission in accord with results for the
HT model [Fig. 2(b)]. At the same time these correlations affect the shape
of the dependence differently for different values of frequency. For
frequencies below the first fundamental resonances, the general shape of the
maximum is not changed, while the split maxima between the first and the
second resonances is replaced by a smooth single maximum. This difference
reflects the fact that a correlation radius of exponential correlations
becomes an additional length scale in the system. Because of this, the
behavior of the transmission as well as other characteristics should be
different for wavelengths greater and smaller than the correlation radius.
For larger wavelengths, the inhomogeneities associated with the exponential
correlations tend to be averaged out and do not affect the system
considerably. For shorter wavelengths, these inhomogeneities become more
important and wash out some features caused by short-range correlations.

Figs. 3(a-d) present the frequency dependence of the Lyapunov coefficient
for different situations shown in Figs. 2. One can see that a strong
increase of the average transmission reflects the corrisponding increase of
the localization length, $\gamma ^{-1}$. It was shown in Ref.\cite
{randomsuperlattice2} that there exist a universal critical value of the
average transmission, $\langle T\rangle _{cr}=0.26$, which separates
localized states from expanded states in systems with a finite length. For
states with $\langle T\rangle <\langle T\rangle _{cr}$ the localization
length is less than the length of the system, and the corresponding states
are localized. In the reverse situation states are extended. One can see
from the results presented that short-range correlations strongly influence
localization properties of states in finite disordered systems. Correlations
of the TM type do not support delocalization while the structure with $p=1$, 
$q=1/2$ allows delocalized states at frequencies inside forbidden bands of
the structure with no correlations. These new localized states are
different, of course, from states at fundamental resonance frequencies
because they do not survive in infinite systems. At the same time these
states contribute considerably to transport properties of finite, yet
macroscopic systems.

\section{Fluctuation properties of the transmission rate and Lyapunov
exponent}

In this section we consider the effect of correlations on fluctuation
properties of the transmission rate and the Lyapunov exponent. Scaling
properties of the distribution of the transmission rate were studied in Ref.
\cite{randomsuperlattice2}. These properties are known to be universal in a
sense that their dependence upon scaling parameter $t=\gamma L$, where $L$
is a length of the system, remains the same for any kind of structure of a
superlattice. The distribution function of the transmission rate, $W(z,t)$,
where $z=1/T$, is determined as follows \cite{Abrikosov}: 
\[
W(z,t)=\frac{2}{\sqrt{\pi t^{3}}}\int_{x_{0}}^{\infty }\frac{x}{\sqrt{\cosh
^{2}x-z}}\exp {[-(t/4+x^{2}/t)]}dx. 
\]
For well localized states with $t\gg 1$ this distribution reduces to the
normal distribution for $\ln T^{-1}$ with a mean value equal to the Lyapunov
exponent and a standard deviation equal to $2\sqrt{t}$ \cite{Abrikosov}.
Transition between extended and localized states was investigated in Ref. 
\cite{randomsuperlattice2}. The authors of Ref. \cite{randomsuperlattice2}
suggested that $t=2$ is the boundary between the extended and localized
regimes since at this point the average localization length becomes equal to
the size of the superlattice. It can be shown, however, that the mean square
fluctuation of the localization length at this point is also equal to the
size of the system. Therefore, the fluctuations of localization length wash
out a distinctive boundary between these two regimes at $t=2$. At the same
time, one can notice that relative fluctuations of the transmission
coefficient show a sharp increase when average transmission becomes
approximately two times smaller than its value at $t=2$ [Fig. 4(b) in Ref.
\cite{randomsuperlattice1,randomsuperlattice2}]. Based upon this
observation, we find that it is interesting to consider the scaling behavior
of this parameter. Its dependence upon the scaling parameter $t$ obtained by
simulations along with the results of the corresponding theoretical
calculations is shown in Fig. 4(a). We would like to point out at a sharp
increase in relative fluctuations of the transmission at $t\approx 5\div 6$.
It can be seen as an increase in the slope of the averaged curve but also as
a drastic increase of scattering of points in the numerical experiment.
Actually, in order to obtain a more or less smooth line in the region $t>5$
we had to increase the number of realizations for averaging from $200$ for
the region $t<5$ to 2000 for $t>5$. Fig. 4(b) presents the same dependence
with smaller number of averaging equal to $200$. At $t>2$ the fluctuations
of localization length become smaller than the system's size, and localized
states begin to contribute more distinctively to such characteristics as the
relative fluctuations of the transmission rate. One can conclude, therefore,
that a sharp change in the behavior of relative fluctuations of transmission
at $t\approx 5$ can be attributed to the transition between extended and
localized regimes in a finite sample.

The universal relations described above do not imply, however, that
localization properties of individual states at different frequencies are
also universal. Below we present results of our study of fluctuation
properties of localization lengths at some characteristics frequencies of
the system. We are primarily interested in a dependence of these properties
upon the correlation characteristics of the system. In order to study this
problem, we first fix the probability $q=1/2$ and consider the dependence of
the Lyapunov parameter $\gamma $ upon the probability $p$. This choice of
parameters allows one to study the influence of the short-range structure in
which exponential correlations are absent. The value $p=0$ leads to
periodical ordering of the layers with the period equal to $2d$, $p=1/2$
describes the system without correlations, and $p=1$ leads to the structure
opposite to the TM model as was explained in the previous section. As
reference frequencies we consider $k=1.45k_0$ and $k=3.9k_0$, where $k_{0}$
corresponds to the vacuum. In the system without correlations, these
frequencies are positioned in the middle of forbidden bands, becoming
resonance frequencies at $p=1$ (see Fig. 2).

Fig. 5 present results of computer simulations of the Lyapunov exponent
versus the probability parameter $p$ along with theoretical curves based
upon Eq. (\ref{averageepsilon}). It is seen that the Lyapunov exponent at
these frequencies demonstrates $qualitatively$ different behavior. The
Lyapunov exponent at $k=1.45k_0$ shows a monotonic decrease with an increase of
the parameter $p$, while at $k=3.9k_0$ it exhibits a nonmonotonic behavior with
the minimum value at approximately $p=1/2$. The difference in behavior
between these frequencies can be understood if one recalls that $p=0$
corresponds to the periodic structure with a period of $2d$. The frequency $
k=3.9k_0$ falls into a transmission band of this periodic structure, therefore
it demonstrates a small Lyapunov exponent when $p$ approaches $0$. At the
same time the frequency $k=1.45k_0$ falls in a forbidden band for the periodic
structure arising at $p=0$, and, hence, their Lyapunov coefficients sharply
increases at $p\rightarrow 0$. When $p$ approaches $1$ both frequencies
belong to resonance regions associated with the resonance transmission from
blocks with doubled thickness of individual layers. Though the structure
with $p=1$ does not lead to exact doubling of all layers, it does favor such
situation causing a decrease of scattering boundaries and consequently
maxima of transmission at these frequencies. Therefore, the Lyapunov
exponent at all frequencies considered decreases when $p$ approaches $1$.

More detailed information about states corresponding to the selected
frequencies can be obtained from Figs. 6(a,b), which presents relative
fluctuations of the Lyapunov exponent, $\Delta\gamma/\gamma$, and relative fluctuations of the transmission rate, $\Delta T/T$, versus the
probability parameter $p$.  Small $\Delta\gamma/\gamma$ and big
$\Delta T/T$   for $k=1.45k_0$ at small values of $p$ reflect strong localization
of the corresponding states. This is exactly what one would expect for the
states arising in a forbidden gap of a nearly periodic structure. It is
interesting to note, however, that an increase of the degree of a disorder
associated with the increase of $p$ does not enhance localization of the
states. One can see from Figs. 6(a,b) that the state at $k=1.45k_0$ becomes
``less'' localized with increasing $p$. The reason for this behavior is that
an increase of $p$ destroys the periodicity of the structure washing out its
forbidden gaps and weakening opportunities for localization. States at other
frequency show almost delocalized behavior for small $p$ since they belong
to a pass band of the periodic structure and become more localized when
traces of periodicity of the structure gradually disappear aas $p$
approaches $1/2$. For $p>1/2$, both frequencies behave in approximately the
same way, since a memory about their different origin is lost in this
situation.

It is interesting to note that results qualitatively similar to those
presented in Fig. 5 were found in Ref.\cite{randommatrix1}, though the
latter paper dealt with a quite different model. The authors of Ref.\cite
{randommatrix1} studied the effect of ``long-range'' exponential
correlations on localization properties of the nearest-neighbor
tight-binding model with the two-state Markov type distribution of site
energies (the HT model). It was found that at the states far enough from the
band edge and band center of the pure system the Lyapunov exponent exhibits
behavior similar to the curve presented by squares in Fig. 5, and states
at the center of the band behave similar to the second line in this figure.
This similarity can be understood if one considers these two models in their
extreme realizations. We have already discussed that the state at $k=1.45_0$
in our model falls into the forbidden band of the periodic structure arising
at $p=0$. The same is valid for the states in the center of the band in Ref.
\cite{randommatrix1} in the case of extreme ``anticorrelation'' between
adjacent values of the site energies. This similar origin causes similar
behavior when the structures change. The second type of the behavior is
associated with states which belong to pass bands of the respective models,
therefore they also demonstrate similar properties. The third type of
behavior of the Lyapunov exponent found in Ref.\cite{randommatrix1}, in
which the Lyapunov exponent monotonically increases along with the Markov
transition probability, does not exist in our model with the parameter $q$
set to be equal to $1/2$. The reason for this is that the second extreme
structure of Ref.\cite{randommatrix1} corresponds to an almost homogeneous
structure, a situation, whichcan not be realized in our model with $q=1/2$.

\section{Conclusion}

In this paper, we carried out a detailed analysis of the effects of
correlations on localization properties of classical waves in random
superlattices. The correlations between different layers of the superlattice
were introduced within the framework of the generalized random Thue-Morse
model. The statistical properties of the model are controlled by two
parameters $p$ and $q$. By changing the values of these parameters we were
able to consider different kinds of random structures including the
classical random Thue-Morse model and the Hendricks-Teller model introduced
in Ref.\cite{correlsuperlat}, structures with weak random deviations from
periodicity, and others. We found that correlations between the constituent
layers strongly affect localization properties of superlattices and can lead
to a great variety of transmission patterns. This property can allow one to
create superlattices with controlled rates of transmission in different
frequency regions.

We pointed out that relative fluctuations of the transmission rate increase
sharply for a value of scaling parameter of $t\approx 5$. This point can be
considered as a more exact threshold between localized and extended states
in finite systems instead of $t=2$ suggested in Ref.\cite
{randomsuperlattice2}.

We also considered the dependence of localization properties of our model
upon the type of short-range structure associated in the model with the
probability parameter $p$. Since knowing the value of the Lyapunov exponent
itself is not enough to determine wether the state considered is localized
or extended, we also considered relative fluctuations of this parameter
along with relative fluctuations of the transmission rate. These quantities
are size independent and, therefore, are convenient for discussing
localization properties. We found that there exist two kinds of states
exhibiting different behavior when $p$ changes from $0$ to $1$. The behavior
of the states is mainly determined by their position in the spectrum of the
deterministic periodic structure arising at $p=0$. The states from pass
bands of this structure show a decrease of their localization length with an
increase of $p$, while states from stop bands depend upon $p$ in
nonmonotonic way. For small values of $p$, the localization length increases
when $p$ increases, reaches its maximum value for $p=1/2$ and for $p>1/2$
their dependence upon $p$ is similar to that of other states of the system.
Comparing these results and to those obtained in Ref.\cite{randommatrix1},
where the tight-binding model with correlations of the Hendricks -Teller
type was considered, shows surprising similarity between them. The general
conclusion that one can draw from this comparison is that the localization
properties of states in 1-$D$ systems depends strongly upon properties of
deterministic systems, which are opposite extremes of the  random systems
considered, and upon the position of the states in the spectra of these
deterministic systems, and the localization properties are less sensitive to
details of the structure of a random system itself.
\section*{Acknowledgments}
We wish to thank A.Z. Genack for useful comments on the
manuscript. We also benefitted from discussions with A.A. Maradudin and A.R. McGurn. This 
work was supported by the NSF under grant No.
DMR-9311605, by a CUNY collaborative grant, and by NATO Science Programm and Cooperation
Partner Linkage Grant HTECH LG 960919.

\pagebreak
\section*{FIGURE CAPTIONS}
Fig. 1. The correlation function,  $K(r)=<\epsilon(0)\epsilon(r)> - <\epsilon^2>$, for different types of random superlattices. The inserts represent the Fourier transforms of the functions $K(r)$. (a) - $p=1$,$q=1/2$;  (b) - $p=1$,$q=0.6$; (c) - $p=q=0.8$
\vskip 1cm

Fig. 2. Average transmission rate for different types of random superlattices. Circles in this figure and in all figures below present the results of computer simulations, and the solid line shows  the theoretical results. (a) - $p=q=0.5$ (model without correlations); (b) - $p=q=0.8$ (HT model); (c) - $p=1,q=0.5$ (short-range correlations only); (d) - Markov TM model, $p=0.5, q=0$, (circles) and  generalized TM model, $p=1, q=0.5$ (squares). The number of layers is equal to $64$; (e) - $p=1,q=0.8$ (short-range and exponential correlations are present).
\vskip 1cm

Fig. 3. Lyapunov exponents for models presented in Figs.2(a-c,d), respectively.
\vskip 1cm

Fig. 4. Relative fluctuations of the transmission rate versus the scaling parameter $t$. Circles present the results of computing, the solid line shows the theoretical results. The numerical data were obtain from averaging over (a) 2000 realizations and (b) 200 realizations.
\vskip 1cm

Fig. 5. The dependence of the Lyapunov coefficient versus the probability parameter $p$ for $q=0.5$. Circles and squares  show results for $k=1.45k_0$ and for  $k=3.9k_0$, respectively; solid lines present corresponding theoretical data.
\vskip 1cm

Fig. 6. Relative fluctuations of the transmission rate (a) and of the Lyapunov coefficient (b) versus the probability parametr $p$. All notations are the same as in Fig. 5.

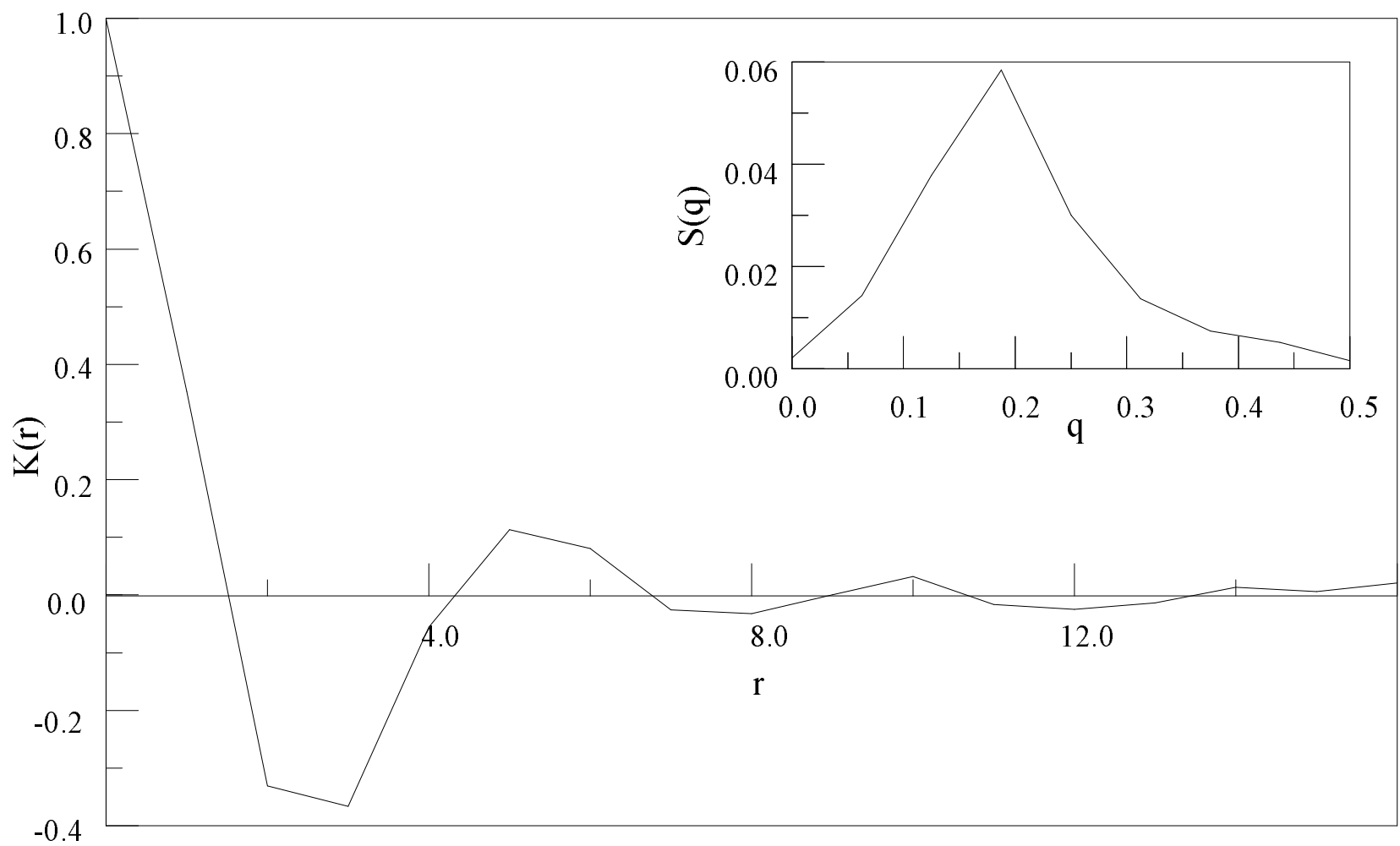
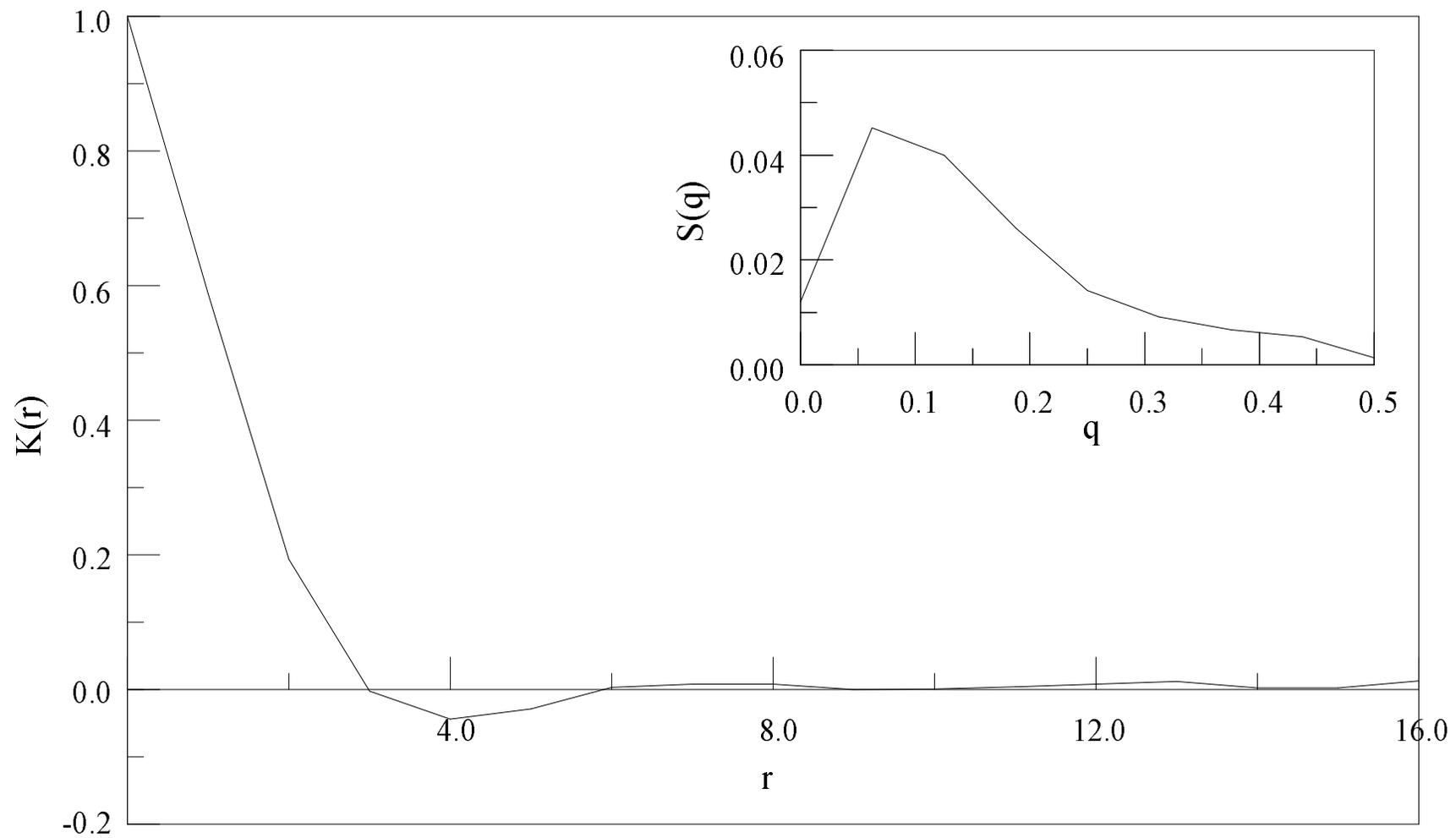
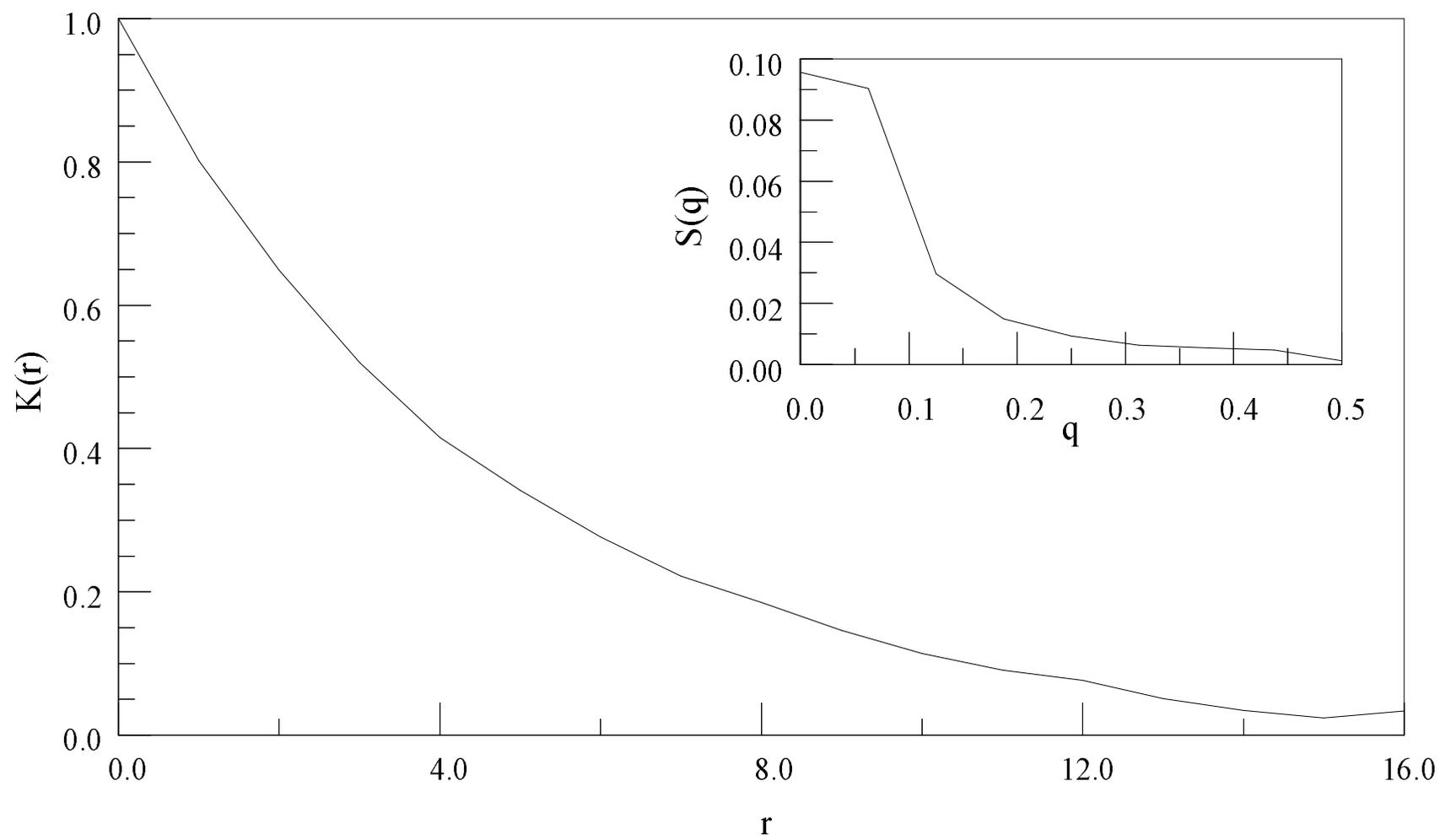
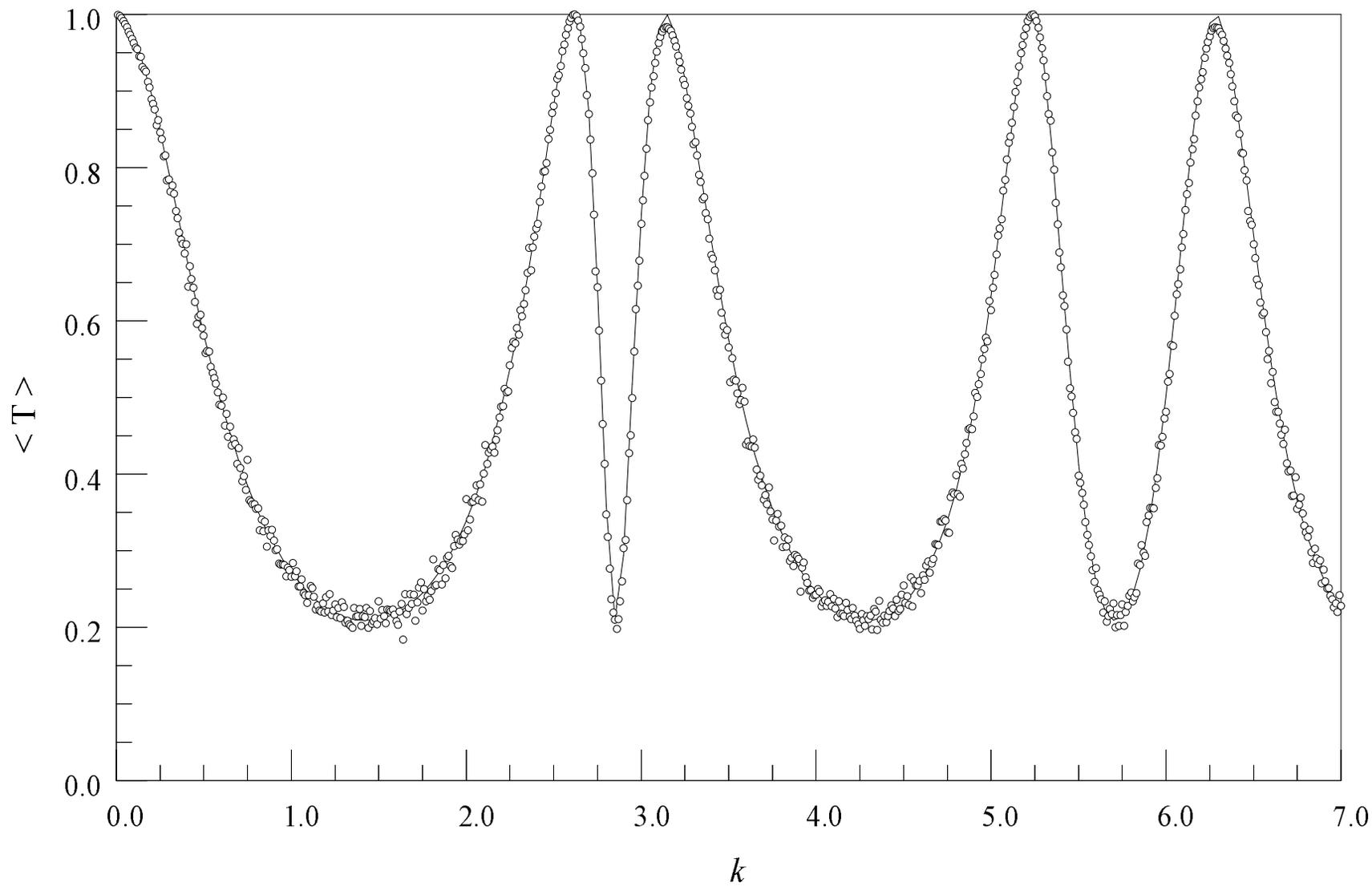
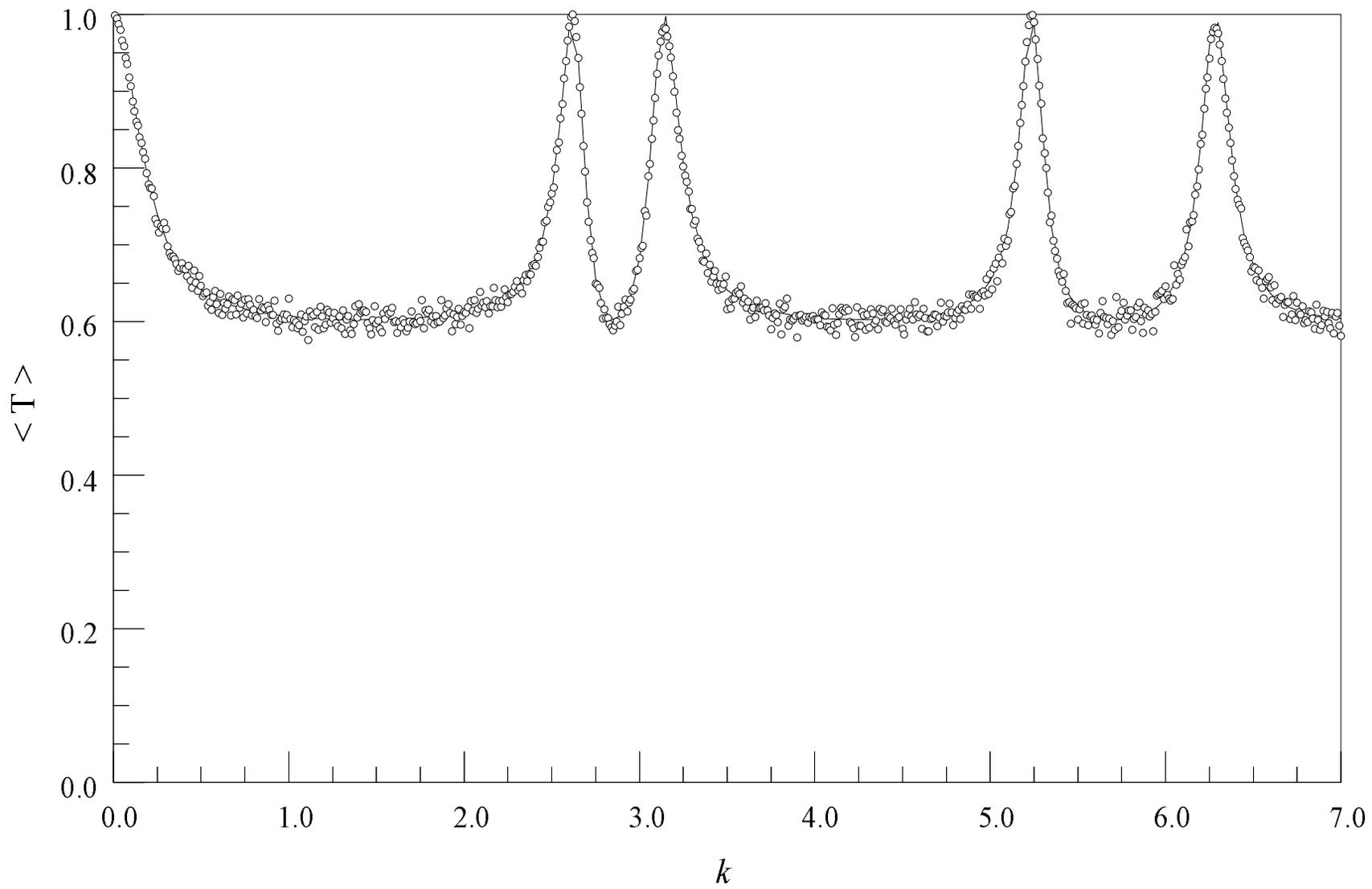
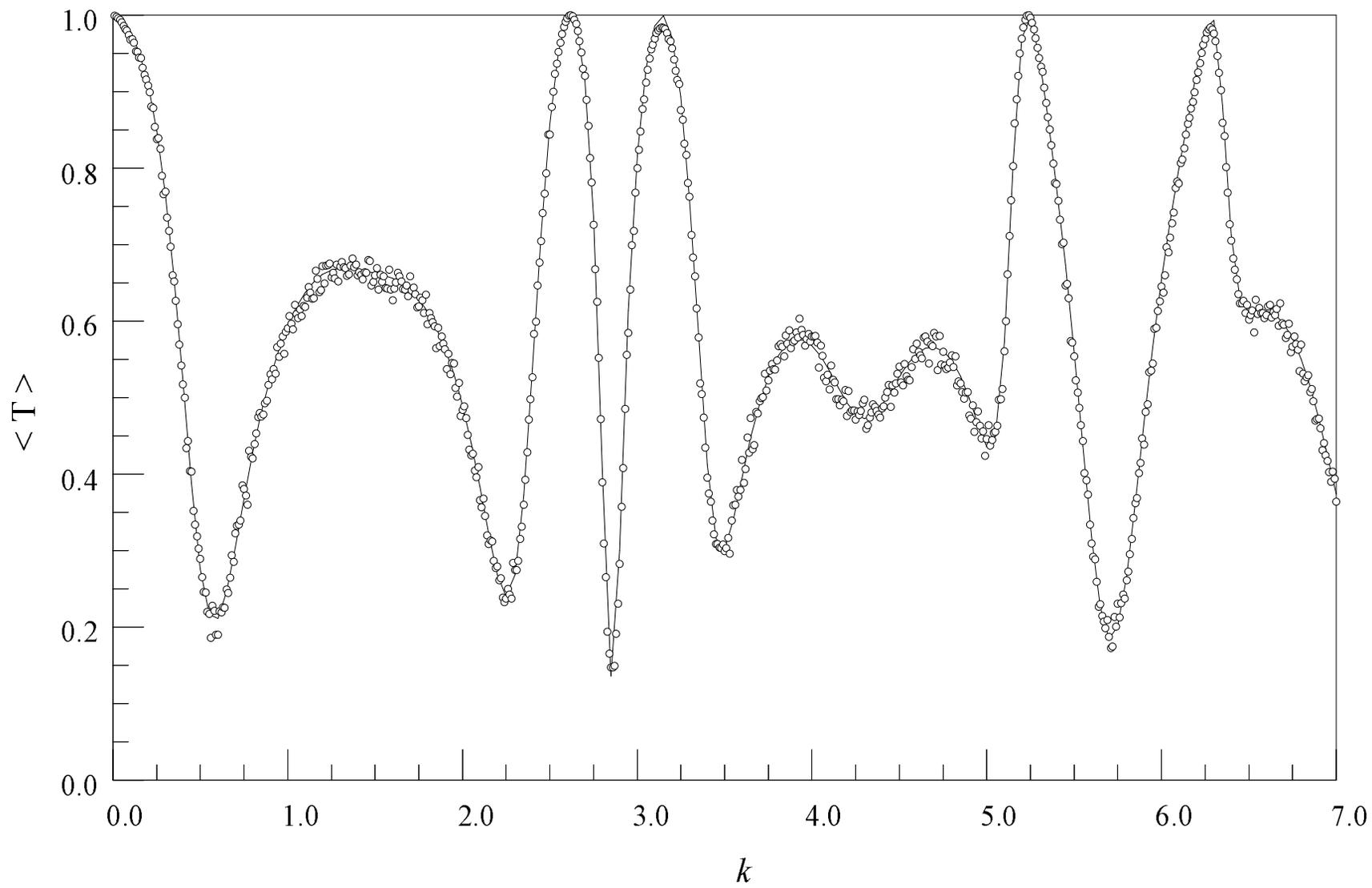
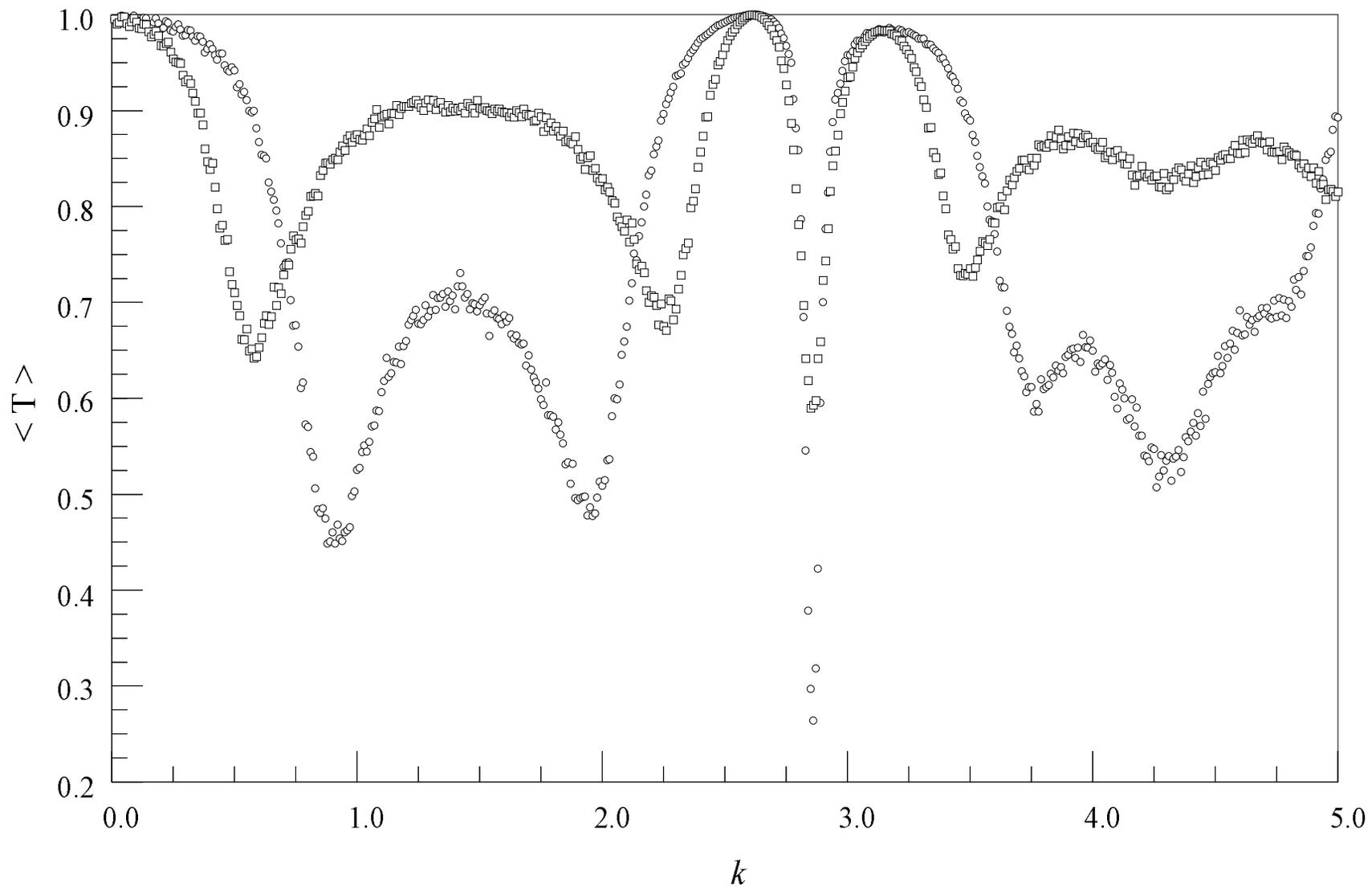
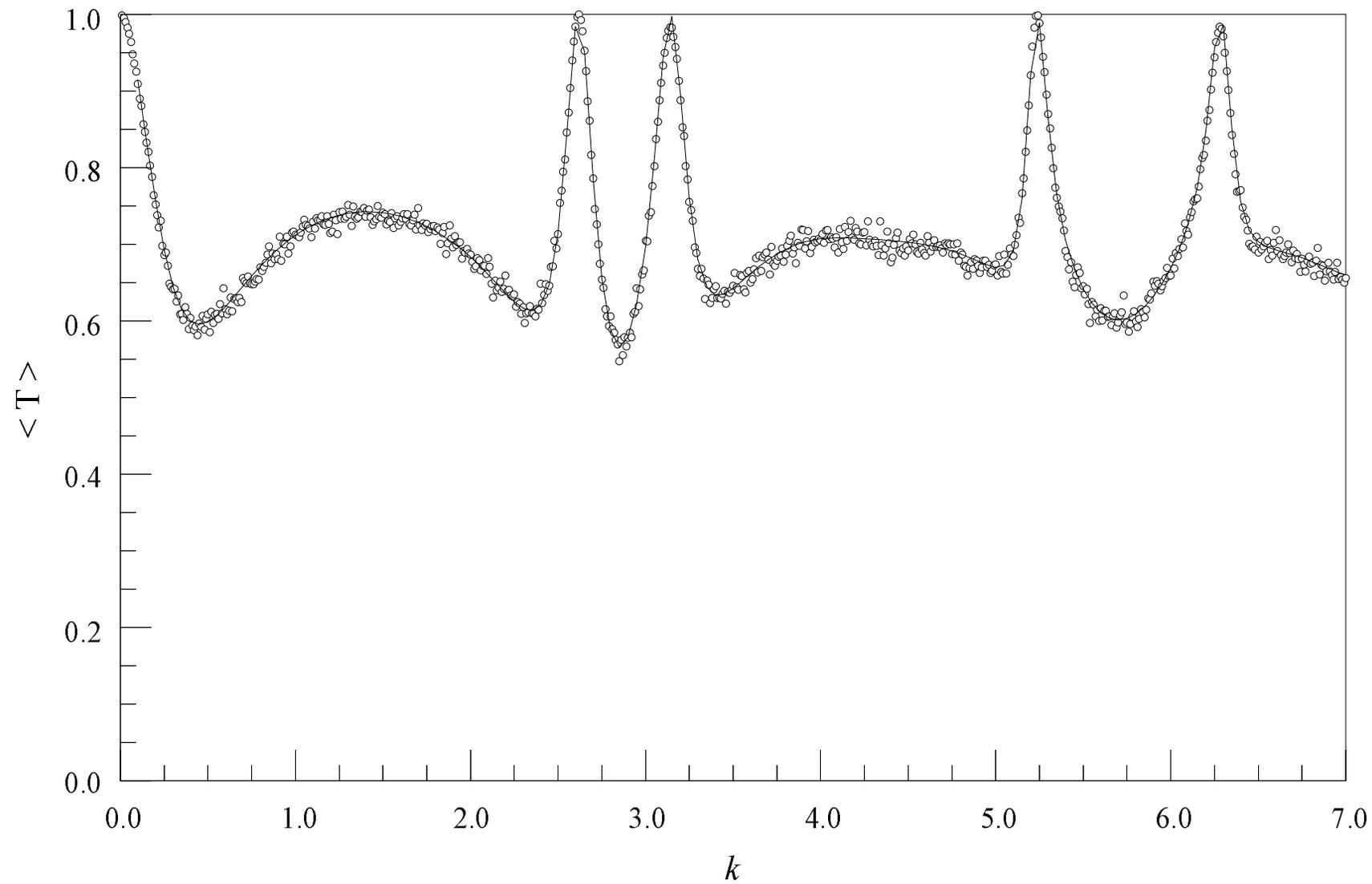
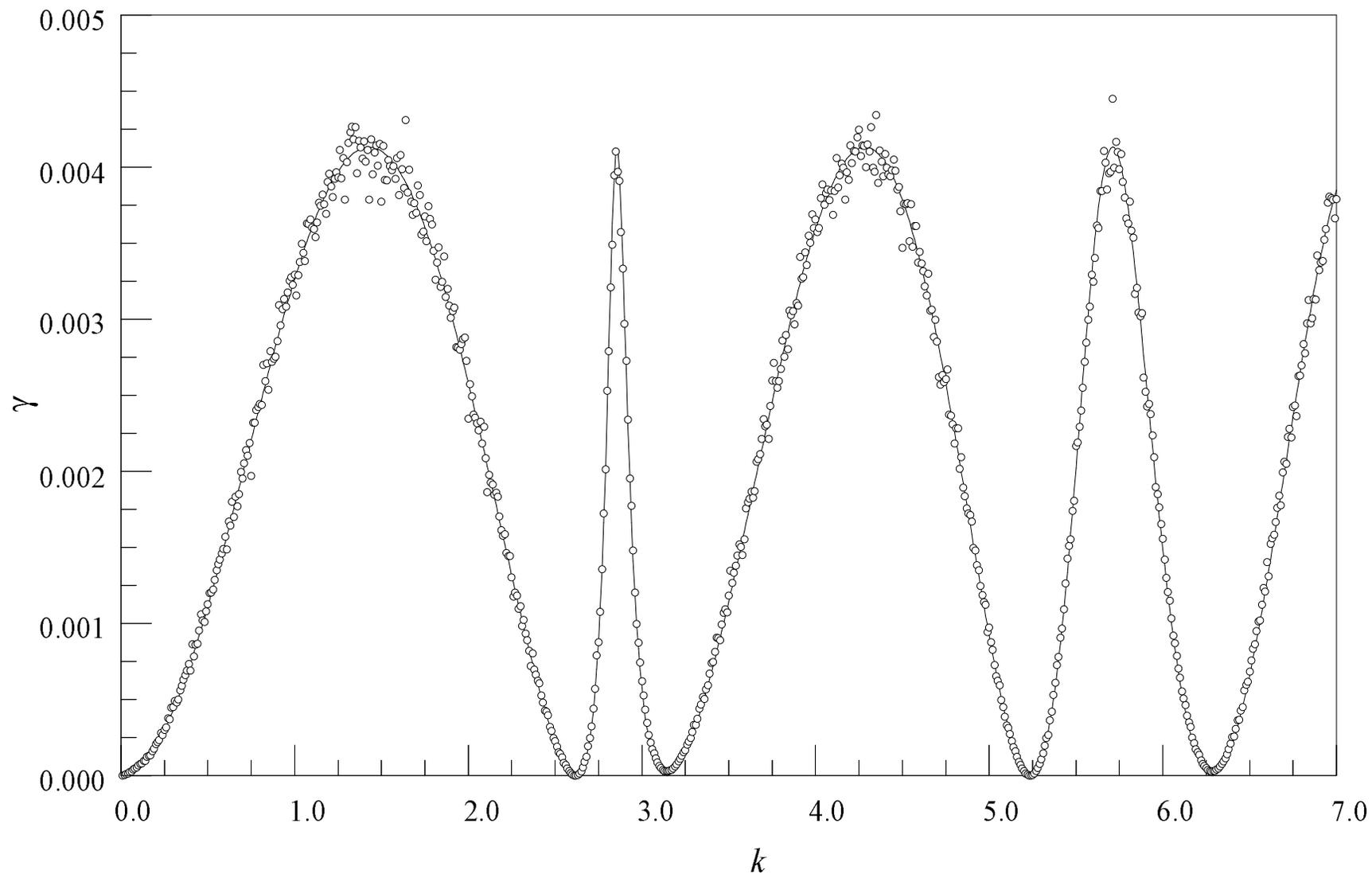
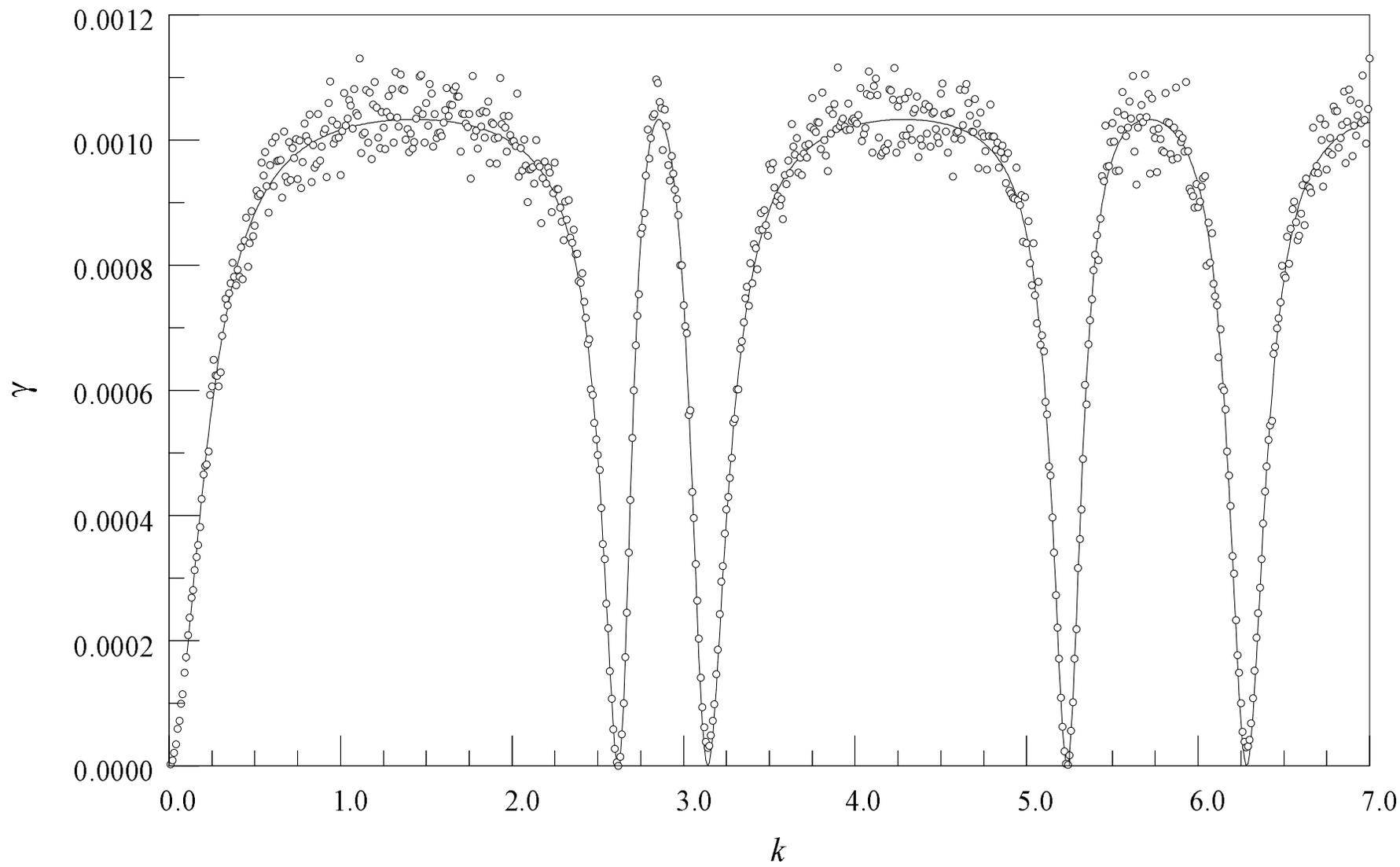
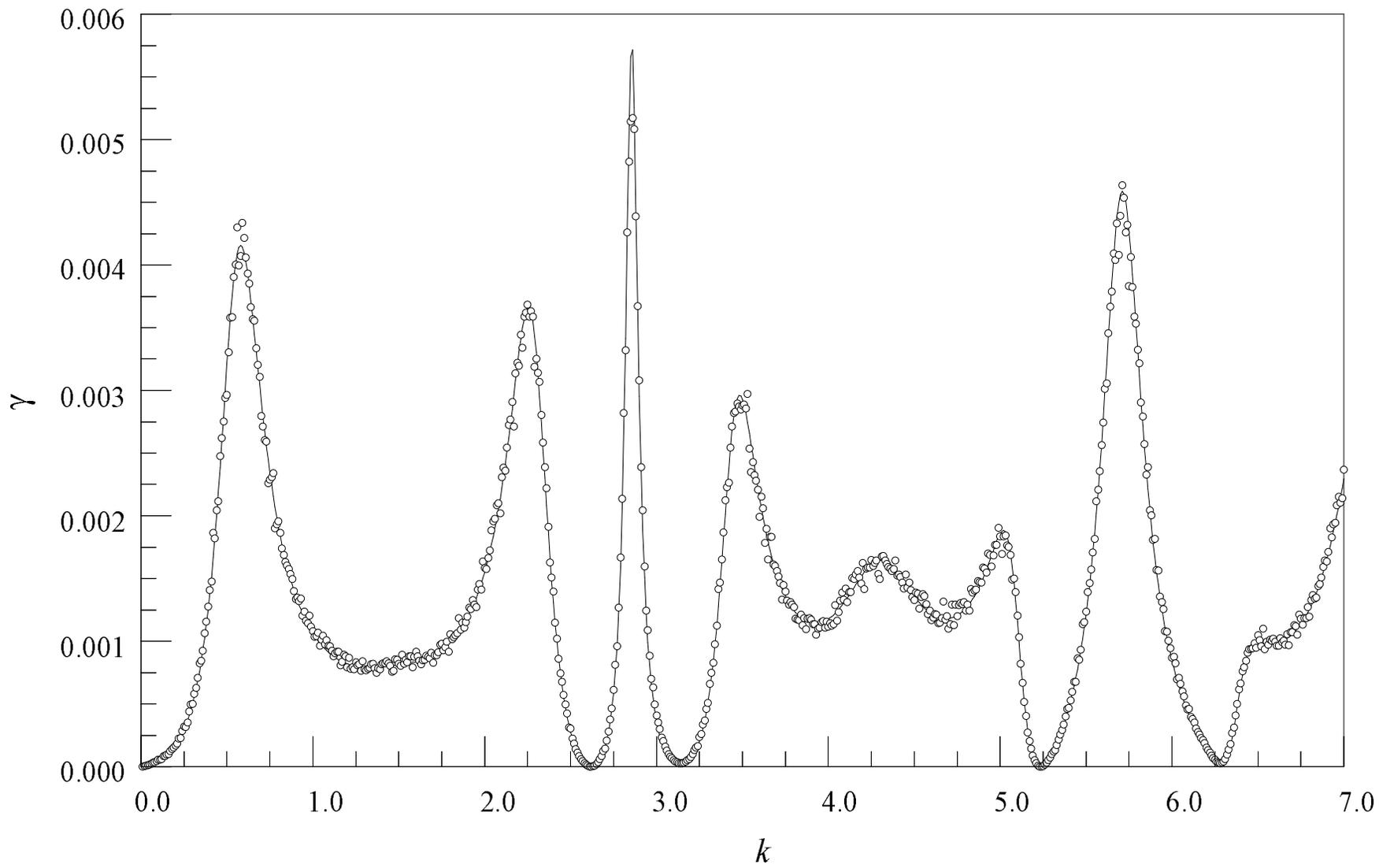
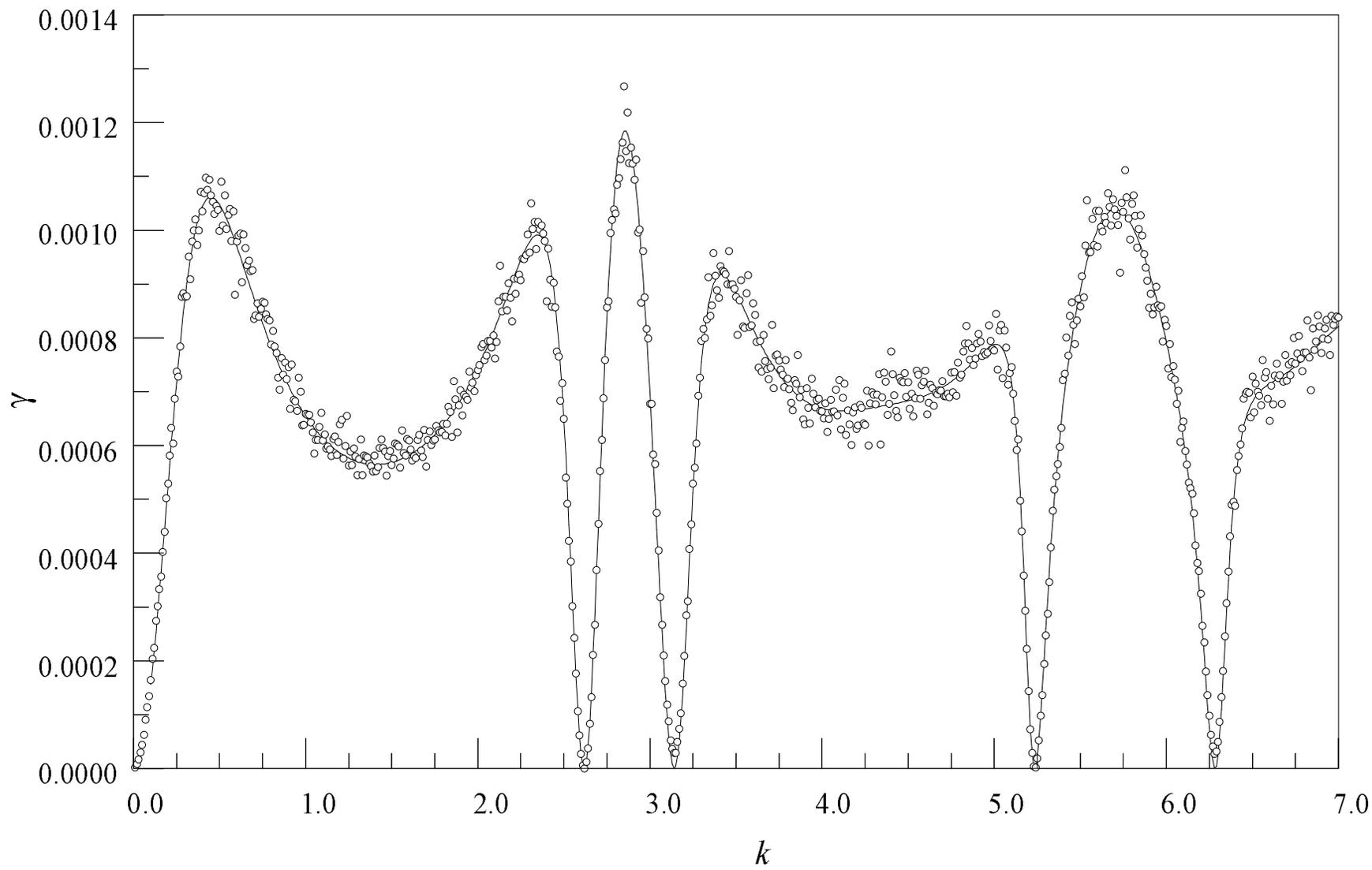
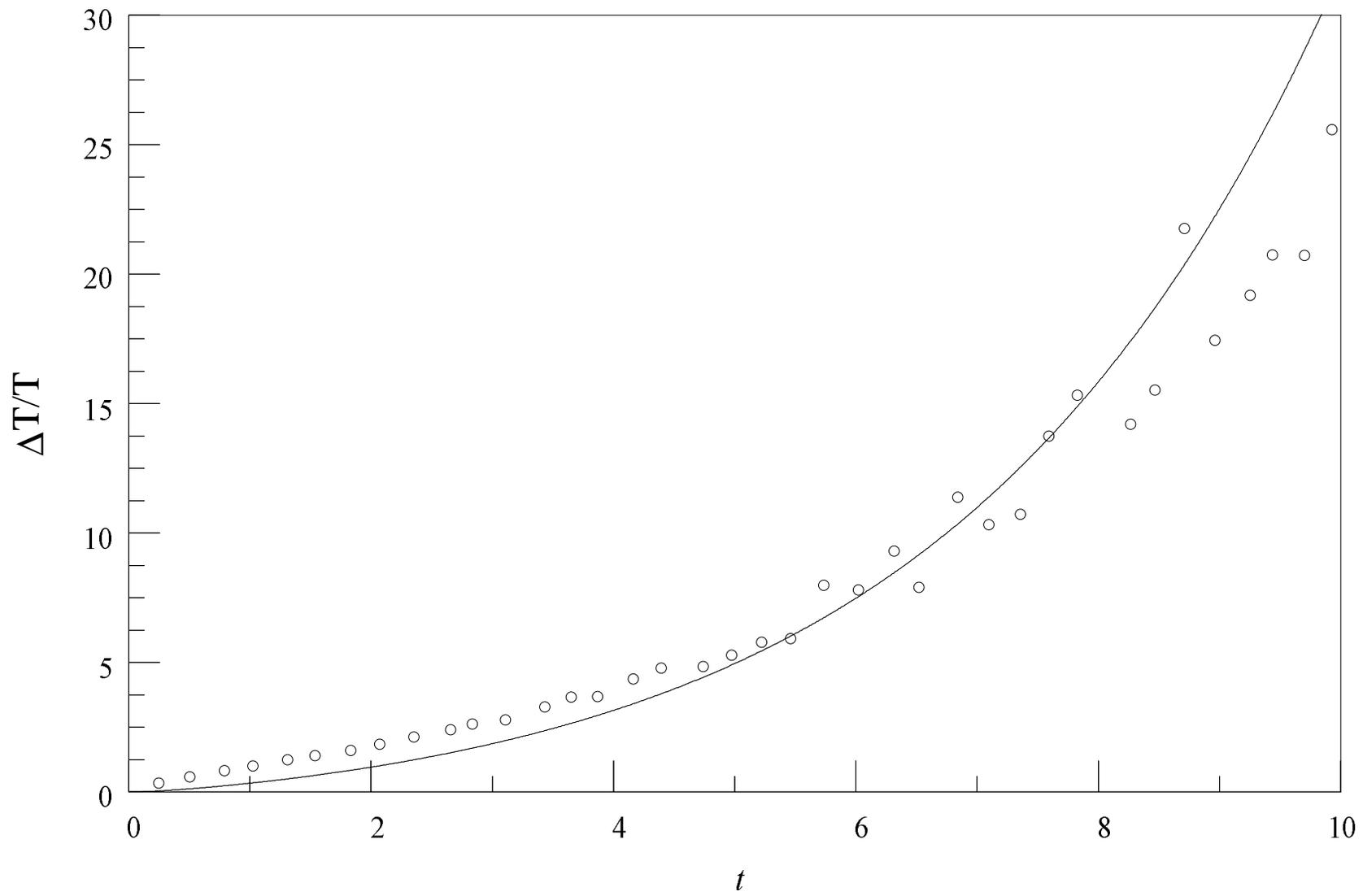
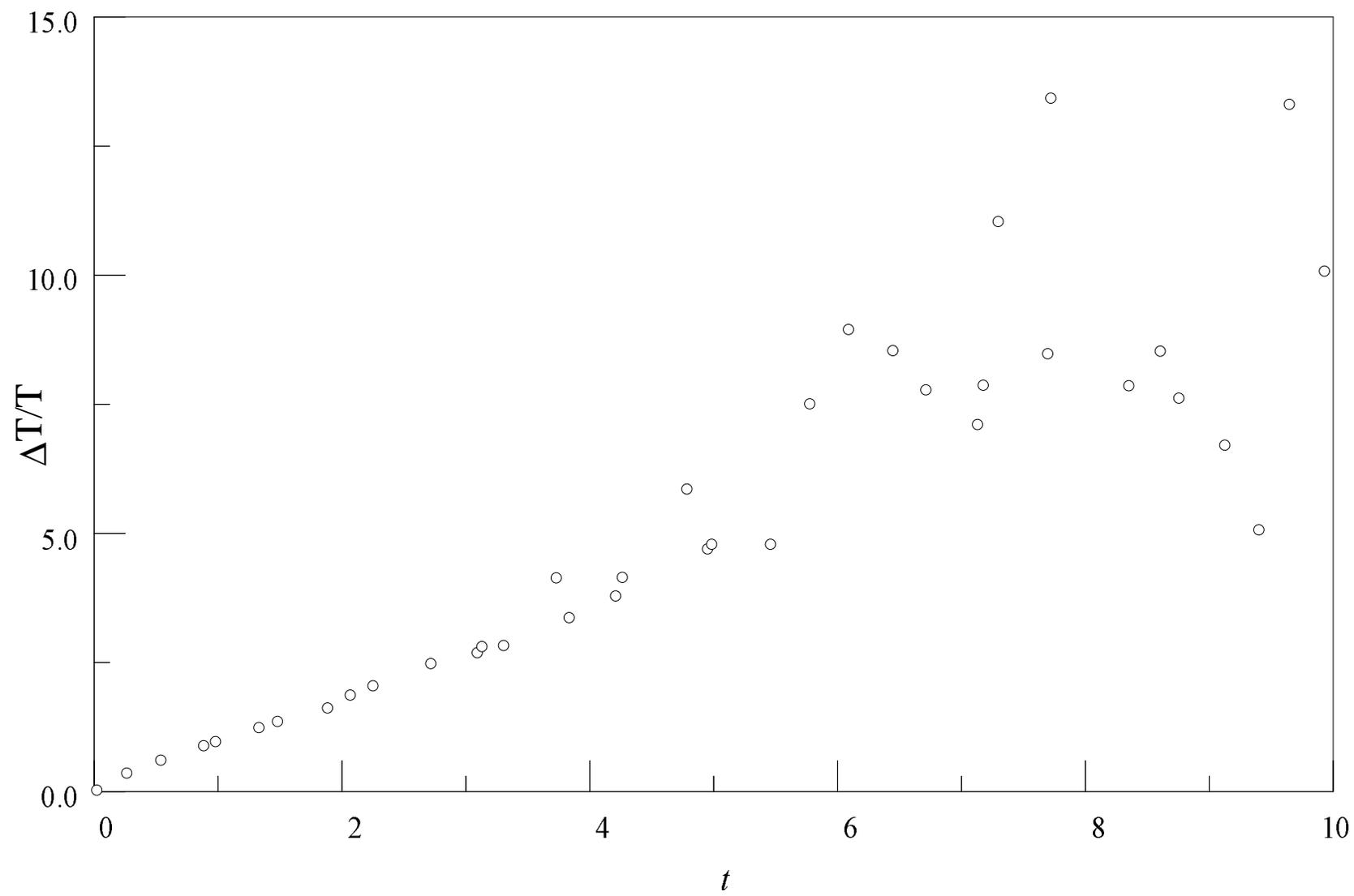
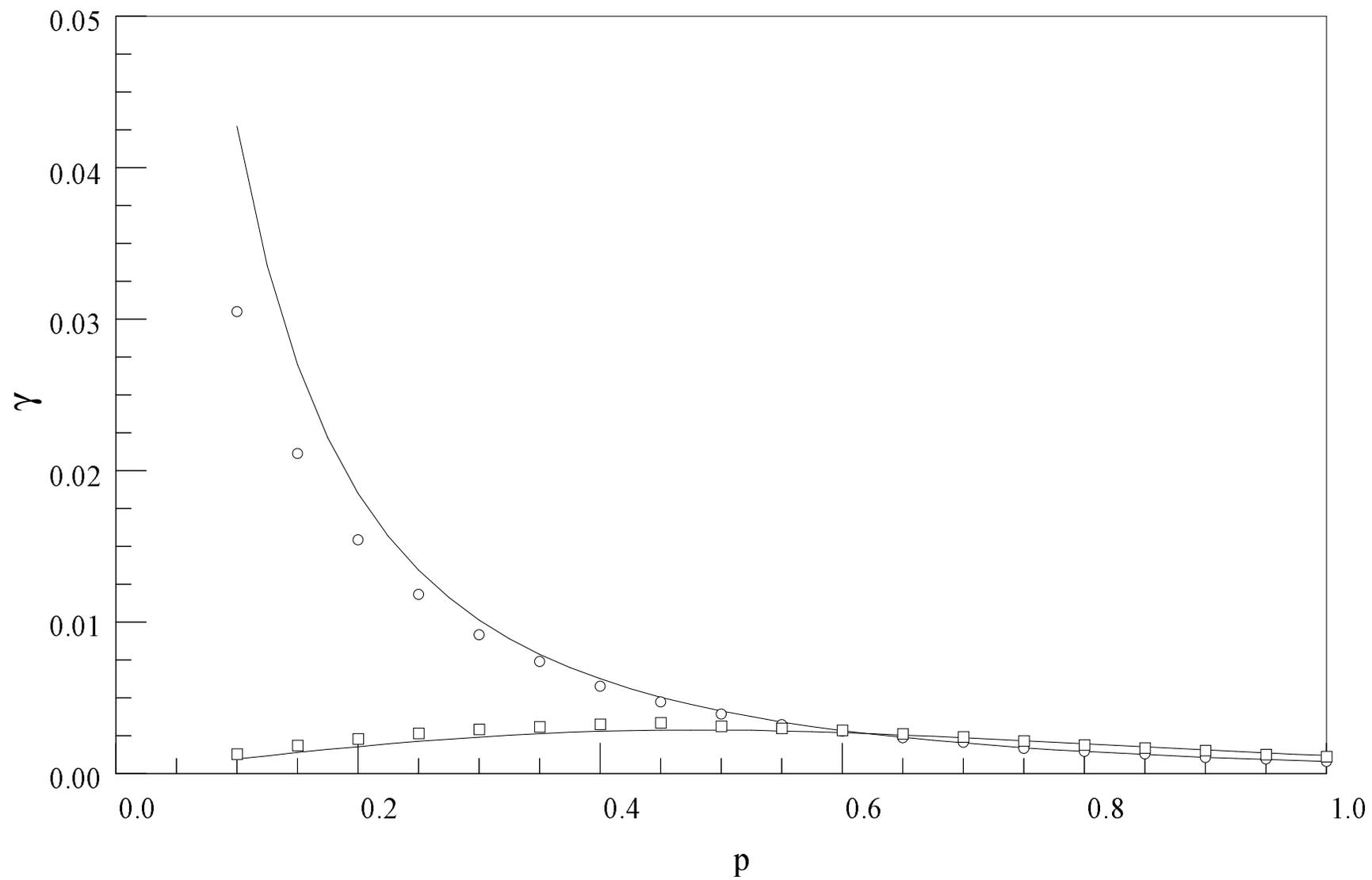
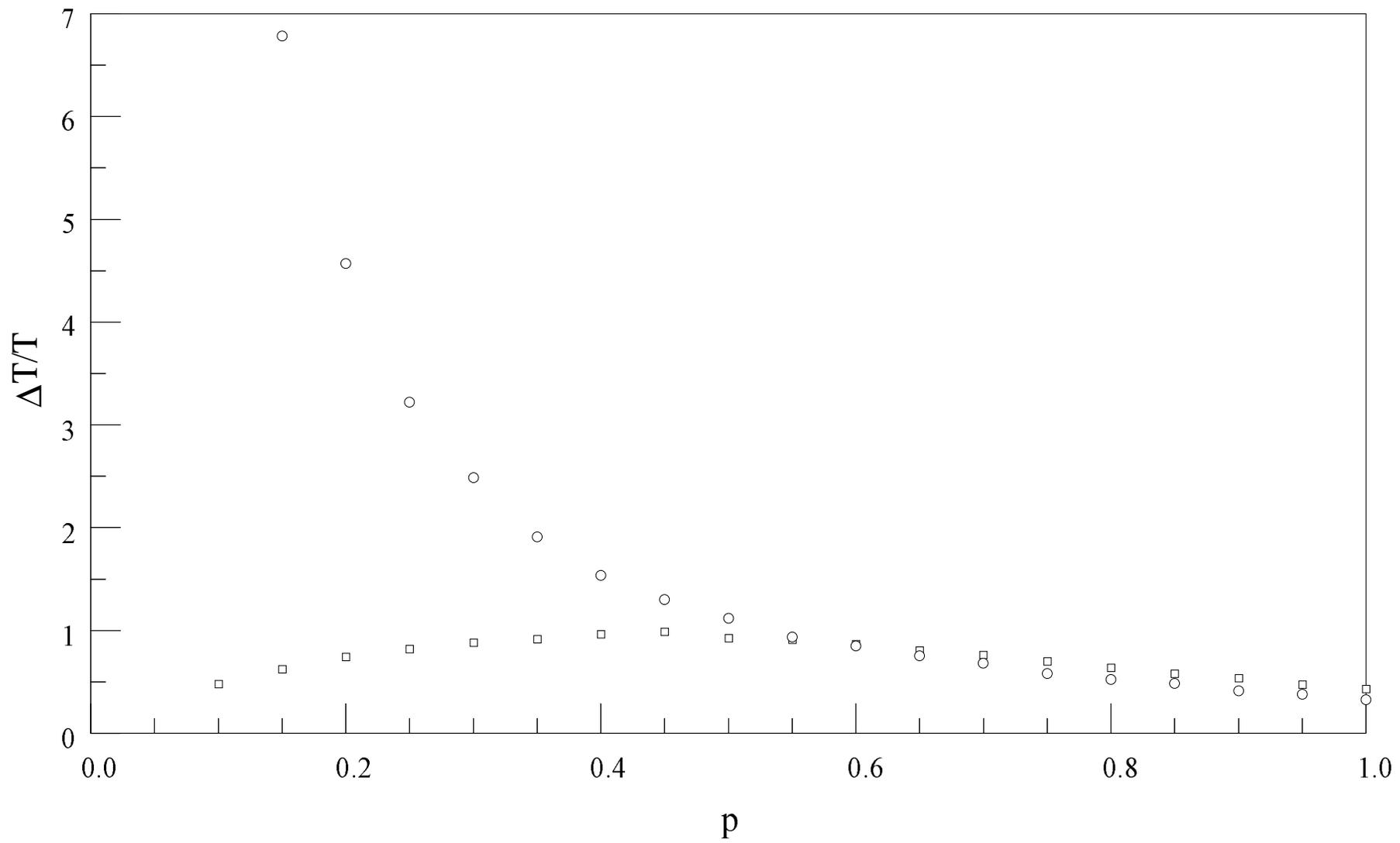
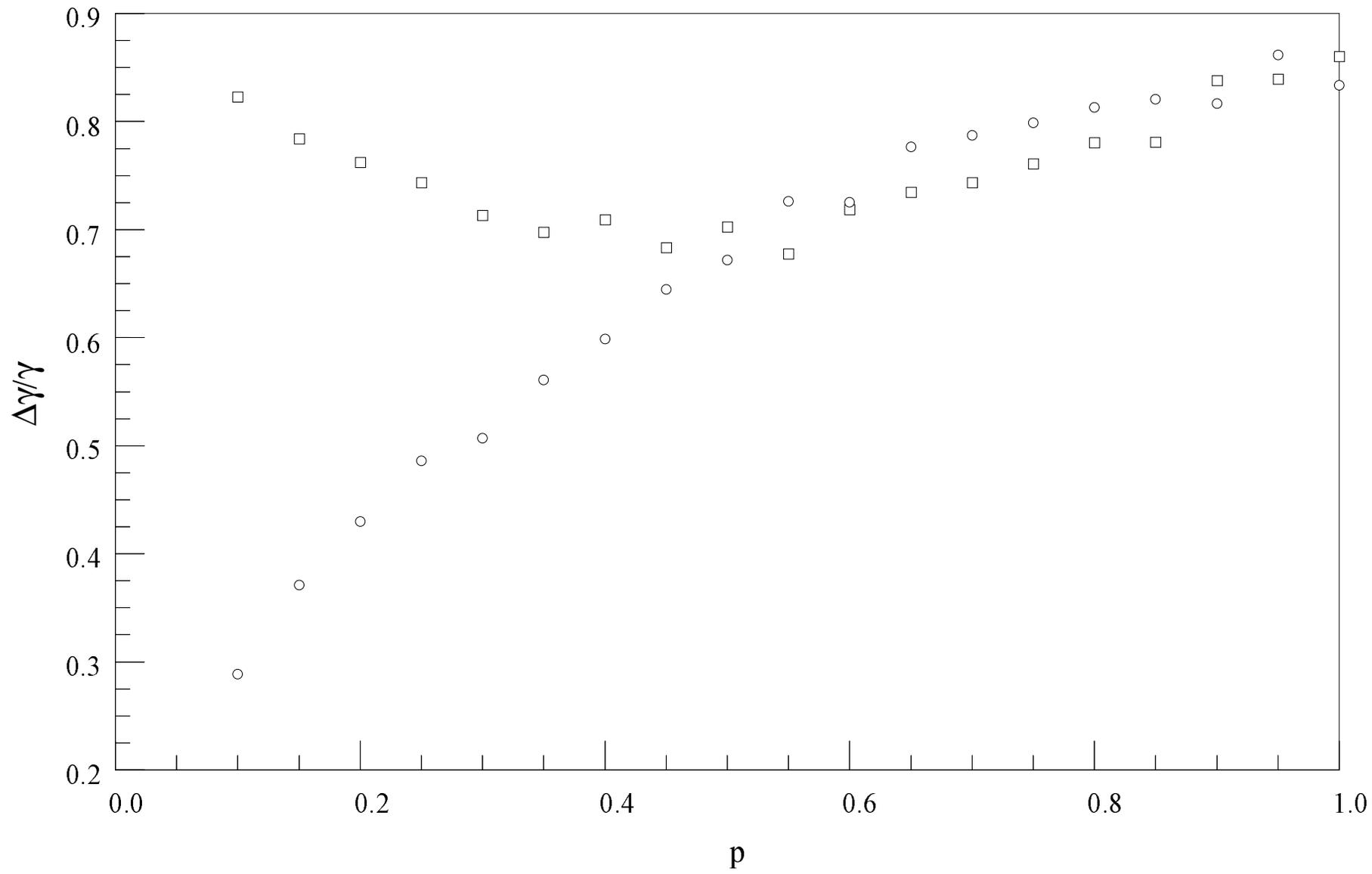
\end{document}